\documentclass[10 pt, conference]{IEEEtran}%
\usepackage{multirow}
\usepackage{subfigure}
\usepackage{threeparttable}
\usepackage{booktabs}
\usepackage{graphicx}
\usepackage{algorithm}
\usepackage{algorithmic}
\usepackage{setspace}
\usepackage{amsmath}
\usepackage{bm}
\usepackage{xcolor}
\usepackage[mathscr]{eucal}
\usepackage{amsfonts}
\usepackage{amssymb}%
\usepackage{amsthm}
\newtheorem*{remark}{Remark}

\begin{document}

\title{Comparison of Cooperative Driving Strategies \\for CAVs at Signal-Free Intersections}
\author{Huile~Xu, Christos G.~Cassandras,~\IEEEmembership{Fellow,~IEEE,}
Li~Li,~\IEEEmembership{Fellow,~IEEE,}
and~Yi~Zhang,~\IEEEmembership{Member,~IEEE}
\thanks{Manuscript received \today; accepted. The work of H. Xu and C. G. Cassandras was supported in part by NSF under grants ECCS-1931600, DMS-166464, CNS-1645681, by AFOSR under grant FA9550-19-1-0158, by ARPA-E¡¯s NEXTCAR program under grant DE-AR0000796 and by the MathWorks.
The work of H. Xu, L. Li, and Y. Zhang was supported in part by the National Key Research and Development Program of China under Grant 2018YFB1600600.}
\thanks{H. Xu is with the Department of
Automation, BNRist, Tsinghua University, Beijing 100084, China and also with
the Division of Systems Engineering and Center for Information and Systems
Engineering, Boston University, Brookline, MA, 02446, USA. (e-mail: hl-xu16@mails.tsinghua.edu.cn)} \thanks{C. G.
Cassandras is with the Division of Systems Engineering and Center for
Information and Systems Engineering, Boston University, Brookline, MA, 02446,
USA. (e-mail: cgc@bu.edu)} \thanks{L. Li is with the Department of Automation, BNRist, Tsinghua
University, Beijing 100084, China. (e-mail: li-li@tsinghua.edu.cn)} \thanks{Y. Zhang is with Department of
Automation, BNRist, Tsinghua University, Beijing 100084, China and also with
the Tsinghua-Berkeley Shenzhen Institute (TBSI), Tower C2, Nanshan
Intelligence Park 1001, Xueyuan Blvd., Nanshan District, Shenzhen 518055,
China (e-mail: zhyi@tsinghua.edu.cn).} }
\maketitle

\begin{abstract}
The properties of cooperative driving strategies for planning and controlling
Connected and Automated Vehicles (CAVs) at intersections range from some that achieve highly efficient coordination performance to others whose
implementation is computationally fast. This paper comprehensively compares
the performance of four representative strategies in terms of travel time,
energy consumption, computation time, and fairness under different conditions,
including the geometric configuration of intersections, asymmetry in traffic
arrival rates, and the relative magnitude of these rates. Our simulation-based
study has led to the following conclusions: 1) the Monte Carlo Tree Search
(MCTS)-based strategy achieves the best traffic efficiency, whereas the
Dynamic Resequencing (DR)-based strategy is energy-optimal; both strategies
perform well in all metrics of interest. If the computation budget is
adequate, the MCTS strategy is recommended; otherwise, the DR strategy is
preferable; 2) An asymmetric intersection has a noticeable impact on the
strategies, whereas the influence of the arrival rates can be neglected. When
the geometric shape is asymmetrical, the modified First-In-First-Out (FIFO)
strategy significantly outperforms the FIFO strategy and works well when the
traffic demand is moderate, but their performances are similar in other
situations; and 3) Improving traffic efficiency sometimes comes at the cost of
fairness, but the DR and MCTS strategies can be adjusted to realize a better
trade-off between various performance metrics by appropriately designing their
objective functions.

\end{abstract}


\IEEEpeerreviewmaketitle

\section{Introduction}


\IEEEPARstart{I}{ntersections} are the main bottlenecks for urban traffic. As
reported in \cite{rios2016survey}, congestion in these areas causes
substantial economic loss to society and significantly increases the travel
time of drivers. Coordination and control problems at intersections are
challenging in terms of safety, traffic efficiency, and energy consumption
\cite{rios2016survey,chen2015cooperative}.

The emergence of Connected and Automated Vehicles (CAVs) is believed to be a
promising way of improving safety, traffic efficiency as well as reducing
energy consumption. With the aid of vehicle-to-vehicle (V2V) and
vehicle-to-infrastructure (V2I) communication, CAVs can obtain real-time
operational data from neighboring CAVs and communicate with the infrastructure
\cite{li2014survey}. These technologies have made it possible to plan better
trajectories for CAVs through optimal control methods, as well as implement
these trajectories in real time.

In recent years, various cooperative autonomous driving strategies have been
proposed to achieve optimal coordination for CAVs driving through signal-free
intersections. The goal of these strategies is to minimize one or several
objectives by scheduling both the crossing order and the control inputs
(speed, acceleration) of all CAVs. Thus, cooperative driving strategies mainly
consist of two parts: 1) a scheduling problem in terms of crossing sequences
and controllable arrival times at conflict areas; and 2) an optimal control
problem in terms of control inputs. This paper focuses on the first problem
and divides the existing strategies into two kinds from the perspective of
crossing sequences, i.e., cooperative driving strategies without resequencing
and cooperative driving strategies with resequencing.

Cooperative driving strategies \emph{without} resequencing mainly refer to the
First-In-First-Out (FIFO) approach where we directly determine the crossing
sequence according to the order of CAVs entering a control zone (defined as an
area around the intersection within which V2I communication is possible). Any
new arriving vehicle does not influence the crossing sequence already
determined for previous CAVs. For example, Stone \emph{et al.} proposed an
autonomous intersection management cooperative driving strategy that divides
the intersection into grids (resources) and assigns these grids to CAVs in a
FIFO manner \cite{dresner2004multiagent,dresner2008multiagent}. Malikopoulos
\emph{et al.} designed a decentralized time-then-energy optimal control
framework for CAVs at intersections where they obtained the desired optimal
arrival times of CAVs based on a FIFO crossing sequence and derived the
energy-optimal analytical solution for controlling CAVs to arrive at the
intersection's conflict (merging) zone at these prescribed arrival times
\cite{malikopoulos2018decentralized}. Zhang and Cassandras further extended
the work by including all possible turns and considering the joint
energy-time-optimal solution \cite{zhang2019decentralized}. In addition, they
incorporated safe distance constraints within the control zone and passenger
comfort within the conflict zone into the trajectory optimization framework.
However, recent studies have shown that the performance of the FIFO crossing
sequence may be far from the optimal solution in at least some cases
\cite{xu2019cooperative}.

Cooperative driving strategies \emph{with} resequencing aim to find a better
crossing sequence for CAVs within the control zone. One of the prevailing
ideas is to formulate an optimization problem whose decision variables are
crossing sequences and control inputs. Specifically, binary variables are
introduced to represent the crossing priority between any two CAVs, which
leads to Mixed-Integer Linear Programming (MILP) problems
\cite{li2017recasting,fayazi2018mixed}. However, MILP problems are NP-hard,
i.e., their computation time increases exponentially with the number of
considered CAVs. Alternatively, it has been shown that the problem may be
treated as a tree search problem where each tree node represents a special
crossing sequence \cite{li2006cooperative}. The equivalent objective is to
find a leaf node that corresponds to the optimal solution. However, this
approach faces similar computational disadvantages as MILP-based strategies.
Although techniques such as grouping methods \cite{xu2018grouping} or pruning
\cite{meng2018analysis} have been proposed to reduce the size of the original
problem or to accelerate the search process, it is still hard to obtain a
real-time solution for complicated driving scenarios that arise, for example,
in multi-lane intersections. To overcome the above shortcomings, there have
been several recent studies on this topic. For example, Xu \emph{et al.}
proposed a Monte Carlo Tree Search (MCTS)-based strategy where they used the
MCTS to guide the search process and determine as many promising crossing
sequences as possible within a limited computation budget
\cite{xu2019cooperative}. By performing comparisons with the results of
exhaustive searching (whenever possible), they demonstrated that the MCTS
strategy can always find a near-optimal solution, even for complicated
multi-lane intersections where the search space is enormous
\cite{xu2019cooperative}. Following a very different approach, Zhang and
Cassandras designed a Dynamic Resequencing (DR) scheme to optimize the
crossing sequence \cite{zhang2018decentralized}. Rather than periodically
replanning crossing sequences for all CAVs as in the MCTS approach, the DR
strategy keeps the original crossing sequence unchanged and updates it only
when a new CAV arrives by inserting it into a suitable position within the
original sequence so as maximally improve performance. Nevertheless, due to
the different models and simulation settings used by different researchers, we
still lack a comprehensive comparative performance evaluation for these
strategies under different driving scenarios.

To analyze the relative advantages and disadvantages of different cooperative
driving strategies, we have selected four representative types of such
strategies: the MCTS strategy \cite{xu2019cooperative}, the DR strategy
\cite{zhang2018decentralized}, the commonly used FIFO strategy, and a modified
FIFO-based strategy. This paper first applies these strategies to a typical
signal-free single-lane intersection with the same arrival rate at each lane
and a symmetric geometrical shape. Then, we vary the length of different lanes
and associated arrival rates in order to investigate the impact of
asymmetrical intersection geometries and asymmetrical arrival rates on these
strategies, respectively. In addition to performance metrics such as travel
time and energy, we also compare the computation time of different strategies
and the number of crossing sequences they have considered during their
computation time. Finally, we discuss the drawback often caused by
resequencing, i.e., unfairness across the different traffic arrival lanes, and
introduce a balancing factor to the DR strategy so as effectively control the
trade-off between fairness and efficiency.

The main contributions of this paper are: 1) to comprehensively evaluate and
compare the performance of representative state-of-the-art cooperative driving
strategies; 2) to analyze the influence of asymmetrical arrival rates and
intersection geometries on these strategies; and 3) to explore the trade-off
between different performance metrics and to propose improvements to existing
strategies based on resequencing.

The paper is organized as follows. \emph{Section II} formulates the optimal
control problem of controlling CAVs passing through a signal-free intersection
safely. \emph{Section III} briefly reviews the four cooperative driving
strategies to be compared. Then, in \emph{Section IV }we conduct a series of
experiments to compare their performance under different simulation settings.
\emph{Section V} discusses the trade-off between fairness and efficiency.
Finally, \emph{Section VI} gives concluding remarks.


\section{Problem Formulation}

\begin{figure}[th]
\centering
\includegraphics[width=8cm]{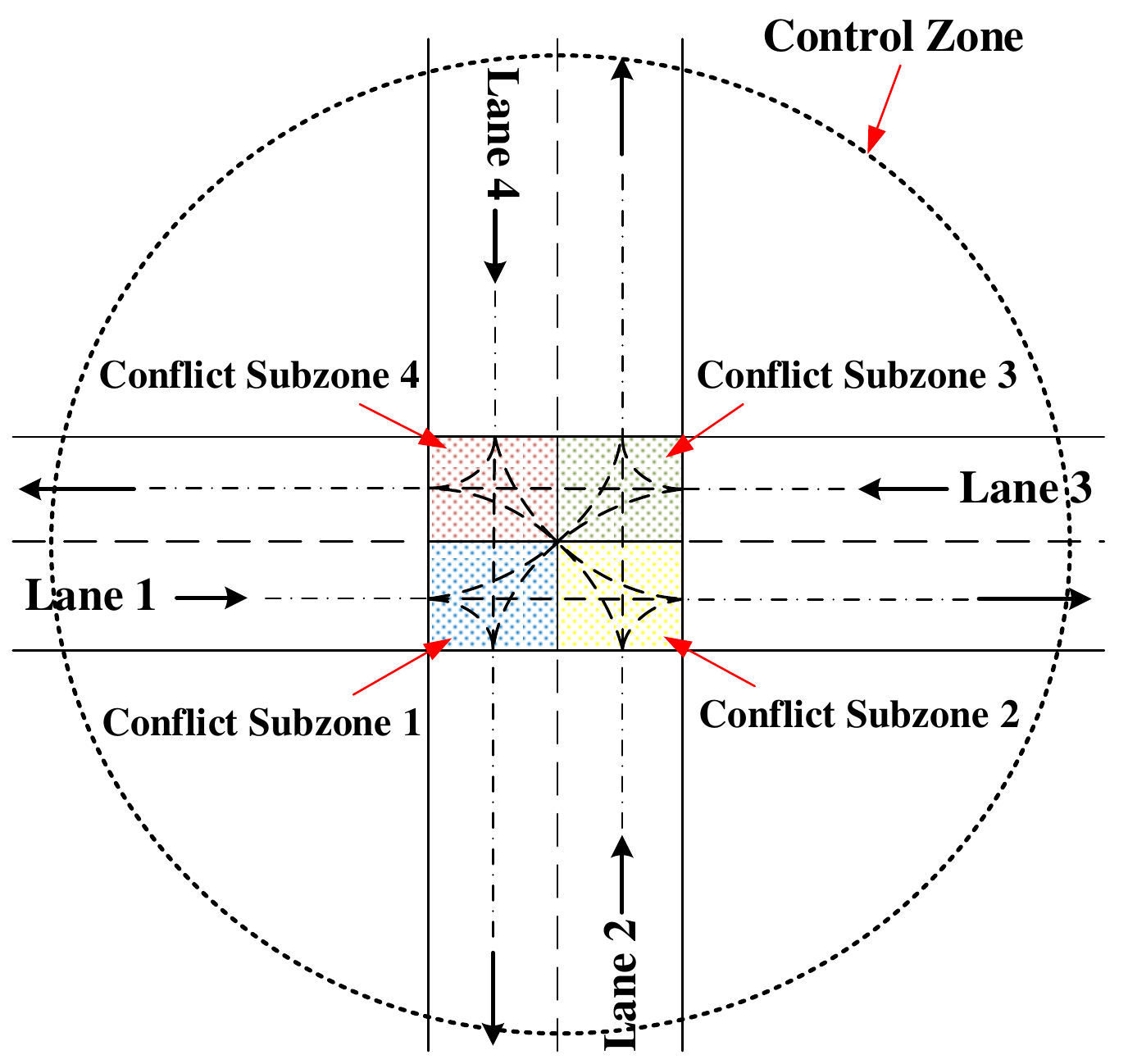}\caption{A typical
intersection.}%
\label{fig:intersection}%
\end{figure}

Fig. \ref{fig:intersection} shows a typical intersection configuration with
a single lane in each direction. The area within the circle is called the
\emph{Control Zone}, while the shadowed area is called the \emph{Conflict
Zone} where lateral collisions may happen. The road segment from the entry of
the Control Zone to the entry of the Conflict Zone is referred to as a control
zone segment, and its length is denoted by $L_{i}$, $i\in\{1,2,3,4\}$. The
value of $L_{i}$ is usually associated with the communication range of
road-side infrastructure equipment (often referred to as a road-side unit). If
all $L_{i}$ are equal, then the intersection is symmetrical; otherwise, it is
an asymmetrical intersection. To improve space utilization, we divide the
Conflict Zone into several subzones. For example, the Conflict Zone in Fig.
\ref{fig:intersection} is divided into 4 Conflict Subzones, which are labeled
Conflict Subzone 1 through Conflict Subzone 4. After this division, CAVs that
pass through different subzones can cross the intersection at the same time.

When a CAV enters the Control Zone, we assign it a unique identity $i$,
labeling it as the $i$th CAV. Then, we use the sequence $Z_{i}$ to denote the
Conflict Subzones that CAV $i$ will pass through. For example, $Z_{i}=\{4,1\}$
in Fig. \ref{fig:intersection} means CAV $i$ will pass through Conflict
Subzone 4 and then Conflict Subzone 1 in sequence.

To simplify the problem, we adopt the following assumptions:

\begin{itemize}
\item Each vehicle instantly shares its complete driving state (position,
velocity, etc.) and intentions with other CAVs via V2V communication (or V2I
then I2V).

\item Similar to \cite{malikopoulos2018decentralized} and
\cite{choi2018reservation}, the velocities of CAVs are constant when passing
through the Conflict Zone.
\end{itemize}

As already mentioned, cooperative driving strategies consist of two parts. We
obtain the optimal crossing sequence and corresponding arrival time by solving
the scheduling problem in the first part. Then, the arrival time is utilized
as the terminal time for solving an optimal control problem in the second
part, through which a CAV's inputs are determined. It is worth noting that
some studies combine these two parts into a single optimization problem
\cite{hult2016coordination}, but the computation time for solving such a
problem is prohibitively large even when tools such as model predictive
control are employed \cite{qian2015decentralized}. Moreover, it is hard to
extend this method to problems with complicated objective functions and
vehicle dynamics. On the other hand, if the crossing sequence is given in
advance, then it is possible to efficiently design a decentralized optimal
control problem for jointly optimizing the arrival times and control inputs of
each CAV as in \cite{zhang2019decentralized} and \cite{xiao2019decentralized}.
We regard these methods as extensions of the basic cooperative coordination
problem and will, therefore, not consider them here.

\subsection{Scheduling problem in terms of arrival times and crossing
sequences}

Let $a_{i,z}$ denote the desired arrival time to the Conflict Subzone $z$ for
CAV $i$, and $\sigma_{i,z}$ is the minimum arrival time to the conflict
subzone $z$ when CAV $i$ travels with the maximum velocity and maximum
acceleration. It is clear that $\sigma_{i,z}$ is the fixed lower bound for
$a_{i,z}$. Let $Z_{i}^{1}$ be the first element in $Z_{i}$, e.g., $Z_{i}%
^{1}=4$ when $Z_{i}=\{4,1\}$. Thus, $a_{i,Z_{i}^{1}}$ is the arrival time at
the first conflict subzone that CAV $i$ will pass through. We also use
$C_{i,z}$ to include the set of indices of all CAVs that may collide with CAV
$i$ in the Conflict Subzone $z$. Once $i$ is known, $Z_{i}^{1}$ and $C_{i,z}$
are fully determined.

We introduce binary variables $\bm{b}=[b_{1,2},b_{1,3},\ldots,b_{n-1,n}]$ to
represent crossing sequences, where $n$ is the number of CAVs currently in the
Control Zone and not yet at the Conflict Zone. We use $b_{i,j}=1$ to indicate
that CAV $i$ is assigned to cross the conflict zone before CAV $j$ for every
$j\in C_{i,z}$, $z\in Z_{i}$, such that $j$ may conflict with $i$; otherwise,
$b_{i,j}=0$ indicating that CAV $j$ has higher crossing priority. Observe that
$(i,j)$ with $j\in C_{i,z}$ is uniquely defined since $j$ can only conflict
with $i$ in a single subzone $z\in Z_{i}$. Therefore, the vector $\bm{b}$
always contains elements which allow us to interpret it as a crossing sequence
in the form of a string. For example, $\bm{b}=[b_{1,2},b_{1,3},b_{2,3}%
]=[1,1,0]$ implies the crossing sequence is $132$. It is also clear that
$b_{i,j}=1-b_{j,i}$, so we omit all $b_{j,i}$ with $j\geq i$ in the definition
of $\bm{b}$.

In addition, let $\bm{a}$ denote the vector of all $a_{i,z}$. We can then
formulate the following optimization problem:%

\begin{subequations}
\begin{align}
&  \min_{\bm{a},\bm{b}}\ \sum_{i=1}^{n}(a_{i,Z_{i}^{1}}-\sigma_{i,Z_{i}^{1}%
})\\
&  \text{s.t.}\ a_{i,z}\geq\sigma_{i,z}\quad i=1,\ldots,n,\text{ }z\in Z_{i}\\
&  a_{i,Z_{i}^{1}}-a_{p,Z_{i}^{1}}\geq\Delta t,\text{ }i=1,\ldots
,n\label{eq1}\\
&  a_{i,z}-a_{j,z}+M\cdot b_{i,j}\geq\Delta t,\text{ }i\in N,\text{ }z\in
Z_{i},\text{ }j\in C_{i,z}\label{eq2}\\
&  a_{j,z}-a_{i,z}+M\cdot(1-b_{i,j})\geq\Delta t,\text{ }i\in N,\text{ }z\in
Z_{i},\text{ }j\in C_{i,z}\label{eq3}\\
&  N=\{1,2,\ldots,n\}\\
&  b_{i,j}\in\{0,1\}
\end{align}
\label{eq:optimization}
\end{subequations}

Constraints (\ref{eq1}) capture the safety rear-end constraints for all CAVs
in the same lane and CAV $p$ is the CAV physically preceding (ahead of) CAV
$i$. Constraints (\ref{eq2}) and (\ref{eq3}) are safety lateral constraints
for CAVs $i$ and $j$ to ensure that there is no more than one vehicle in any
Conflict Subzone at any time. $\Delta t$ is the safety time headway between
two CAVs. $M$ is a sufficiently large number such that if $b_{i,j}=1$, then
inequality (\ref{eq2}) must hold (due to the large value of $M$). It follows
that inequality (\ref{eq3}) takes on the same form as (\ref{eq1}). Thus, if
$b_{i,j}=1$, CAV $i$ is prioritized to cross the Conflict Zone earlier than
CAV $j$.

\begin{remark}
Regarding the selection of a value for $M$, it is straightforward to derive a
finite lower bound for it such that any value greater than this lower bound
may be chosen. In particular, let $a_{\max,z}$ be the arrival time at Conflict
Subzone $z$ when a CAV starts at a Control Zone entry point with the minimum
initial velocity. Then, for any $a_{i,z}$ we have $a_{i,z}=\max(a_{k,z}+\Delta
t,\sigma_{k,z})$ where $k$ is the last CAV passing through Conflict Subzone
$z$ prior to CAV $i$. Since, $\sigma_{k,z}\leq a_{\max,z}$, we have
$a_{i,z}\leq a_{\max,z}+(n-1)\Delta t$. Then, since $a_{i,z}>0$,
\[
a_{j,z}-a_{i,z}+\Delta t<a_{\max,z}+(n-1)\Delta t+\Delta
t<a_{\max,z}+N\Delta t
\]
where $N$ is the capacity of the intersection in terms of the number of CAVs
it can accommodate.
\end{remark}

By solving the optimization problem (\ref{eq:optimization}) to obtain a
solution $(\bm{a},\bm{b})$, we get the optimal crossing sequence (given by
$\bm{b}^{\ast}$) and the desired arrival times for all CAVs. However,
(\ref{eq:optimization}) is a MILP problem whose computation time increases
exponentially with the number of CAVs.

To indirectly solve the problem, \cite{malikopoulos2018decentralized} and
\cite{xu2018grouping} pointed out that we can determine the crossing sequence
first and then the primal problem reduces to a linear programming problem that
can be easily solved. For example, \cite{xu2018grouping} proposed a simple
iterative structure algorithm to derive the desired arrival times for all CAVs
with a time complexity $O(n)$. Based on this idea, the original problem is
transformed into a problem of finding the optimal crossing sequence for
improving traffic efficiency. In recent years, there have been many
state-of-the-art studies on this topic, which will be introduced in detail in
the next section.

\subsection{Optimal control problem in terms of control inputs}

After determining the desired arrival times, we need to plan control inputs
for optimally controlling CAVs so that they arrive at the Conflict Zone at the
desired time and at the same time minimize a specific objective. Aside from
traffic efficiency, energy consumption is a performance metric of interest.
Since the energy consumption of CAV $i$ is a function of its control inputs
and monotonically increasing with the acceleration $u_{i}$, we formulate the
following optimal control problem solved by each CAV\ $i$ in a decentralized fashion:%

\begin{subequations}
\begin{align}
&  \min_{u_{i}(t)}\ \int_{t_{i}^{0}}^{a_{i,Z_{i}^{1}}}\mathcal{C}%
(u_{i}(t))dt\\
\text{s.t.}\  &  \dot{x}_{i}(t)=v(t),\quad\dot{v}_{i}(t)=u(t)\label{eq:22}\\
&  x_{i}(t_{i}^{0})=0,\quad v_{i}(t_{i}^{0})=v_{i}^{0}\label{eq:23}\\
&  x_{i}(a_{i,Z_{i}^{1}})=L\label{eq:24}\\
&  v_{i}(a_{i,Z_{i}^{1}})=v_{i}^{f}\label{eq:25}\\
&  x_{p}(t) - x_{i}(t) \geq l\label{eq:28}\\
&  v_{\min,i}\leq v_{i}(t)\leq v_{\max,i}\label{eq:26}\\
&  a_{\min,i}\leq u_{i}(t)\leq a_{\max,i} \label{eq:27}%
\end{align}
\label{eq:optimization2}
\end{subequations}

\noindent where $\mathcal{C}(\cdot)$ is a strictly increasing function of its
argument, e.g., $\mathcal{C}(u_{i}(t))=\frac{1}{2}u_{i}^{2}(t)$. Constraint
(\ref{eq:22}) consists of the vehicle dynamics where $x_{i}(t)$ and $v_{i}(t)$
are the position and velocity of CAV $i$ at time $t$. Constraints
(\ref{eq:23}), (\ref{eq:24}) and (\ref{eq:25}) are boundary conditions where
$t_{i}^{0}$ is the time instant when CAV $i$ enters the Control Zone,
$v_{i}^{0}$ is the initial speed of CAV $i$, $L$ is the length of the Control
Zone segment, and $v_{i}^{f}$ is the final speed of CAV $i$. Similar to
\cite{malikopoulos2018decentralized,xu2019cooperative}, we assume that the
final speeds of all CAVs are the same and fixed, but this assumption can be
easily relaxed as shown in \cite{zhang2019decentralized} and will not
influence our analysis on crossing sequences. Constraints (\ref{eq:28}) is the
safety rear-end distance constraint where CAV $p$ is the CAV physically ahead
of CAV $i$ and $l$ is the safety distance. Finally, (\ref{eq:26}) and
(\ref{eq:27}) are physical limitation constraints where $v_{\min,i}$ and
$v_{\max,i}$ are the minimum and maximum velocity for CAV $i$, $a_{\min,i}$
and $a_{\max,i}$ are the minimum and maximum acceleration for CAV $i$,
respectively. For simplicity, we assume that $v_{\min,i}$, $v_{\max,i}$,
$a_{\min,i}$, and $a_{\max,i}$ are the same for all CAVs, and we can handle
the situation where these values are dependent on CAV $i$ in the same way.

It is still time-consuming to solve problem (\ref{eq:optimization2}) through
interior point methods or commercial software, e.g., CPLEX. However, due to
its simple vehicle dynamics and constraints, we can derive analytical
solutions for this problem \cite{malikopoulos2018decentralized,
zhang2019decentralized} and quickly obtain the optimal control inputs. It is
worth noting that even if the vehicle dynamics and constraints become
complicated, we can invoke the Control Barrier Function (CBF) methodology to
solve the corresponding optimal control problem efficiently as a sequence of
quadratic problems over discretized time instants in $[t_{i}^{0}%
,a_{i,Z_{i}^{1}}]$. Interested readers are referred to
\cite{xiao2019decentralized}.

\section{Cooperative Driving Strategies}

In this section, we briefly review four cooperative driving strategies used to
determine the crossing sequence.

\subsection{FIFO strategy and modified FIFO strategy}

In the FIFO strategy, the crossing sequence follows the FIFO principle. The
CAV that enters the Control Zone earlier has a higher crossing priority when a
potential conflict with another CAV arises. It is easy to implement this
strategy: we only need to add a new incoming CAV at the end of the original
crossing sequence and remove from it every CAV that has crossed the Conflict Zone.

However, \cite{zhang2018decentralized} found that this strategy may lead to
poor scheduling and possible congestion when the shape of the control zone is
asymmetrical. We propose a simple idea to overcome this problem, i.e., assign
a higher crossing priority to a CAV that is closer to the Conflict Zone. In
other words, all CAVs in the Control Zone calculate their distance to the
Conflict Zone, and the crossing sequence is derived by sorting CAVs in
ascending order in terms of this distance. We call this new strategy the
\textquotedblleft modified FIFO strategy\textquotedblright\ and implement it
in a time-driven way, i.e., the crossing sequence is periodically updated.

The FIFO strategy and modified FIFO strategy only consider one possible
crossing sequence according to their corresponding defining rule. It is easy
to see that their time complexities are $O(n)$ and $O(nlog(n))$, respectively,
where $n$ is the number of CAVs in the Control Zone. The FIFO strategy is
\emph{event-driven}, since it is only invoked whenever a CAV enters the
Control Zone or leaves the Conflict Zone so as to update the crossing
sequence. The modified FIFO strategy is \emph{time-driven}, since the crossing
order is periodically updated based on the current distance of CAVs from the
Conflict Zone; in particular, the crossing sequence is updated every $T$ seconds.

\subsection{Dynamic Resequencing (DR) strategy}

An improvement over strategies based on a single possible crossing sequence is
to evaluate several feasible crossing sequences whenever a new CAV enters the
Control Zone and to select the optimal one. This is referred to as Dynamic
Resequencing. This strategy maintains the relative order of the remaining CAVs
and finds an appropriate position in which the new CAV can be inserted so as
to optimize a given objective function $\mathcal{J}$. The DR process is shown
in \textbf{Algorithm 1}. Observe that the DR strategy is implemented in an
\emph{event-driven} way with \textbf{Algorithm 1} invoked only when the
triggering event (a new CAV entering the Control Zone) occurs.

\begin{algorithm}
\caption{DR-based cooperative driving strategy}
\begin{algorithmic}[1]
\begin{spacing}{1.2}
\REQUIRE The original crossing sequence $\mathcal{S}$ and the information of all CAVs
\ENSURE A new crossing sequence $\mathcal{S}_{new}$
\STATE Find the preceding vehicle of the new vehicle and its position $k$ in $\mathcal{S}$
\FOR {each $i$ = length($\mathcal{S}$) : $-1$ : $k$}
\STATE Insert the new vehicle into the position $i+1$ of $\mathcal{S}$ and obtain a feasible crossing sequence $\mathcal{S}_{f}$.
\STATE
Compute the corresponding objective value $\mathcal{J}_{f}$ for $\mathcal{S}_{f}$ \label{objectivevalue}
\IF {$\mathcal{J}_{f} < \mathcal{J}_{optimal}$}
\STATE    $\mathcal{J}_{optimal} = \mathcal{J}_{f}$
\STATE    $\mathcal{S}_{new} = \mathcal{S}_{f}$
\ENDIF
\ENDFOR
\RETURN $\mathcal{S}_{new}$
\end{spacing}
\end{algorithmic}
\end{algorithm}

Since the time complexity of computing the objective value of one crossing
sequence is $O(n)$ (see step \ref{objectivevalue} in \textbf{Algorithm 1}),
the worst time complexity of DR strategy is $O(n^{2})$. However, the expected
computational complexity is actually $O(Mn)$ where $M$ is the number of lanes.
A proof and analysis of the DR strategy and its complexity can be found in
\cite{zhang2018decentralized}.

\subsection{Monte Carlo Tree Search (MCTS) strategy}

As mentioned above, the DR strategy keeps the original crossing order of CAVs
unchanged and determines an appropriate insertion position for any new
arriving CAV. In contrast, the MCTS-based strategy aims to find the globally
optimal crossing sequence among all feasible crossing sequences at every time
instant. Clearly, for a real-time implementation it is difficult to enumerate
all feasible solutions within a limited computation time, especially when the
number of CAVs is large. Thus, this strategy combines a MCTS with some
heuristic rules for guiding the search process so as to traverse as many
promising crossing sequences as possible.

\begin{algorithm}
\caption{MCTS-based cooperative driving strategy}
\begin{algorithmic}[1]
\begin{spacing}{1.2}
\REQUIRE The information of all CAVs
\ENSURE A crossing sequence $\mathcal{S}_{best}$
\STATE Initialize a root node.
\WHILE{the computation budget is not reached}
\STATE Selection: starting at the root node, select the most urgent expandable node based on the UCB1 policy \cite{kocsis2006bandit}.
\STATE Expansion: randomly select one unvisited child node of the most urgent expandable node to be a new node that is added to the tree.
\STATE Simulation: run several rollout simulations to determine a complete crossing sequence based on the partial crossing sequence represented by the current new node to evaluate the potential of the new node. Some heuristic rules are utilized to help us quickly capture the real potential of a node during simulation. If the objective value of the crossing sequence obtained from simulation is better than the currently optimal value, record it in $\mathcal{S}_{best}$.
\STATE Backpropagation: the simulation result is backpropagated through the selected nodes to update the scores of all its parent nodes.
\ENDWHILE
\RETURN $\mathcal{S}_{best}$
\end{spacing}
\end{algorithmic}
\end{algorithm}

\textbf{Algorithm 2} outlines the idea of the MCTS-based strategy; interested
readers are referred to \cite{xu2019cooperative,browne2012survey}. Similar to
the modified FIFO strategy, the MCTS is also \emph{time-driven} with the
crossing sequence updated every $T$ seconds.

If there is no computation time limit imposed, the time complexity of the MCTS
strategy is exponential $O(2^{n})$, which is highly undesirable. Nevertheless,
the maximum computation time we set can ensure that the search process is
finished within an acceptable time dictated by the specific scenario, e.g.,
100 $ms$. As validated in \cite{xu2019cooperative}, the MCTS combined with
heuristic rules can always lead to a near-optimal or the optimal crossing
sequence even when the search is limited to a very small subset of the search space.

\section{Simulation Results}

In this section, we conduct a series of simulations to compare the performance
of the four different strategies outlined in Section III. We assume that
vehicle arrivals occur according to Poisson processes (four different ones and
one for each entry point) and vary the rate parameters of these Poisson
processes to test the performance of the strategies under different traffic
demands. In addition, we vary the values of the Control Zone segment lengths
to investigate the impact of intersection asymmetries on these strategies.

To accurately describe the arrival of CAVs, we adopt the point-queue model in
our simulations \cite{ban2012continuous}. The model assumes vehicles travel in
the free-flow state until they get to the boundary of the intersection we
study. To be more concrete, each lane is associated with an independent
point-queue. Then, for each lane, we generate the same random number of CAVs
generated from a Poisson distribution and let them enter into the point-queue.
If the preceding CAV allows adequate space for the first CAV in the
point-queue, then this CAV will dequeue and enter the intersection Control
Zone. Otherwise, it will stay in the virtual point-queue. In this manner, the
actual arrival process of CAVs at each entry point preserves the feasibility
constraints (\ref{eq:28}) at time $t_{i}^{0}$. Thus, the point-queue model has
the same effect as the feasibility enforcement zone mechanism described in
\cite{zhang2017optimal}.

In the following comparison, if a strategy is implemented in a time-driven
way, we update the crossing sequence every 2 seconds. For each scenario, we
simulate a 20-minute traffic process to decrease the influence of random
factors. The maximum computation budget for the MCTS strategy is set as 100
$ms$, i.e., the outcome of \textbf{Algorithm 2} is used after its execution
time reaches this value, unless it has already terminated.

Our performance comparison over the four different strategies is based on two
indicators, travel time (delay) and energy consumption.

\begin{enumerate}
\item The travel time (delay) of CAV $i$ is defined as
\begin{equation}
d_{i}=a_{i}-\sigma_{i},
\end{equation}

\noindent where $a_{i}$ is the arrival time at the Conflict Zone for CAV $i$,
and $\sigma_{i}$ is the minimum arrival time at the Conflict Zone when CAV $i$
travels at its maximum velocity and acceleration.

\item The energy consumption of CAV $i$ is defined as%

\begin{equation}
E_{i} = \int_{0}^{a_{i}} u_{i}^{2}(t)dt,
\end{equation}

\noindent where $u_{i}(t)$ is the control input of vehicle $i$ at time $t$. In
actuality, $E_{i}$ above is only an approximation of a vehicle's energy
consumption, since such consumption also depends on speed and deceleration
does not normally contribute to it. Thus, we can replace the expression above by a more detailed energy model as in \cite{kamal2012model}. Although we have performed such additional computations, we have found that they do not provide any significant new insights relative to the results shown in the following.




\end{enumerate}

\subsection{Comparison results for intersection with same arrival rates and
symmetrical geometry}

In this experiment, we set the lengths of all control zone segments to be 250
$m$ and the arrival rates at the entry of all lanes at the same value. Then,
we vary the arrival rates from 90 veh/hour/lane to 420 veh/hour/lane to test
the performance of the four cooperative driving strategies under different
traffic demands. The results are shown in Table \ref{tab:tablel}.

\begin{table*}[ptbh]
\caption{The comparison results of different strategies for the symmetrical
intersection}%
\label{tab:tablel}
\centering
\begin{threeparttable}
\begin{tabular}{cccc}
\toprule
arrival rates (veh/h/lane) & strategies & average delay ($s$) & average energy consumption \\
\midrule
\multirow{4}[2]{*}{90} & MCTS  & 2.1770 & 0.0253 \\
& DR    & 2.1770 & 0.0253 \\
& modified FIFO & 2.2353 & 0.0399 \\
& FIFO  & 2.2353 & 0.0399 \\
\midrule
\multirow{4}[2]{*}{180} & MCTS  & 2.4447 & 0.0975 \\
& DR    & 2.4634 & 0.0849 \\
& modified FIFO & 2.5822 & 0.1077 \\
& FIFO  & 2.5822 & 0.1077 \\
\midrule
\multirow{4}[2]{*}{270} & MCTS  & 2.6552 & 0.2186 \\
& DR    & 2.7270 & 0.1376 \\
& modified FIFO & 3.1263 & 0.2586 \\
& FIFO  & 3.1263 & 0.2586 \\
\midrule
\multirow{4}[2]{*}{360} & MCTS  & 3.2535 & 0.6379 \\
& DR    & 3.4248 & 0.3578 \\
& modified FIFO & 4.4207 & 0.6980 \\
& FIFO  & 4.4145 & 0.6973 \\
\midrule
\multirow{4}[2]{*}{450} & MCTS  & 4.0706 & 0.9212 \\
& DR    & 4.4325 & 0.7664 \\
& modified FIFO & 6.6261 & 2.0094 \\
& FIFO  & 6.3254 & 1.8600 \\
\bottomrule
\end{tabular}%
\begin{tablenotes}
\footnotesize
\item[1] The arrival rates at the entries of all lanes are the same.
\item[2] The lengths of all lanes are the same.
\end{tablenotes}
\end{threeparttable}
\end{table*}

\begin{table*}[ptbh]
\caption{The comparison results of different strategies for the geometrically
asymmetrical intersection}%
\label{tab:table2}
\centering
\begin{threeparttable}
\begin{tabular}{cccc}
\toprule
arrival rates (veh/h/lane) & strategies & average delay ($s$) & average energy consumption  \\
\midrule
\multirow{4}[2]{*}{90} & MCTS  & 2.1934 & 0.0724 \\
& DR    & 2.1934 & 0.0724 \\
& modified FIFO & 2.2276 & 0.0950 \\
& FIFO  & 3.0406 & 1.0353 \\
\midrule
\multirow{4}[2]{*}{180} & MCTS  & 2.4491 & 0.1925 \\
& DR    & 2.4592 & 0.1817 \\
& modified FIFO & 2.6281 & 0.2582 \\
& FIFO  & 3.9139 & 1.8821 \\
\midrule
\multirow{4}[2]{*}{270} & MCTS  & 2.6569 & 0.3697 \\
& DR    & 2.7529 & 0.4375 \\
& modified FIFO & 3.0658 & 0.5392 \\
& FIFO  & 4.6047 & 2.7749 \\
\midrule
\multirow{4}[2]{*}{360} & MCTS  & 3.2322 & 0.9428 \\
& DR    & 3.4250 & 0.9020 \\
& modified FIFO & 4.5203 & 1.1992 \\
& FIFO  & 5.9534 & 3.7580 \\
\midrule
\multirow{4}[2]{*}{450} & MCTS  & 3.8585 & 2.3008 \\
& DR    & 4.2526 & 1.5560 \\
& modified FIFO & 6.4420 & 2.3650 \\
& FIFO  & 8.5438 & 5.5766 \\
\bottomrule
\end{tabular}%
\begin{tablenotes}
\footnotesize
\item[1] The arrival rates at the entries of all lanes are the same.
\item[2] The length of one lane is 150 $m$, while the lengths of remaining lanes are 250 $m$.
\end{tablenotes}
\end{threeparttable}
\end{table*}

It is clear that the MCTS strategy realizes the best traffic efficiency since
the average delay is the smallest under all traffic demands, while the DR
strategy is best in terms of energy efficiency. This reveals a natural
trade-off between traffic efficiency and energy efficiency. Improving traffic
efficiency requires re-planning the crossing sequence and changing the states
of CAVs, but this always comes at the cost of additional energy consumption
because such re-planning involves frequent acceleration adjustments.

As for the FIFO strategy, one would expect that its energy consumption should
be the least since it keeps the order unchanged and CAVs travel according to
the control inputs planned when they enter the Control Zone. However, due to
the poor coordination performance of the FIFO strategy, especially when the
arrival rate is high, there are always many more CAVs in the Control Zone at
relatively slow speeds. As a result, CAVs usually need to more frequently
decelerate and accelerate compared to other strategies whose coordination
performance is better, which leads to much higher energy consumption than the
MCTS and DR strategies when running a long simulation.

\begin{figure}[th]
\centering
\subfigure[Trajectories from the FIFO-based strategy.]{\includegraphics[width=8cm]{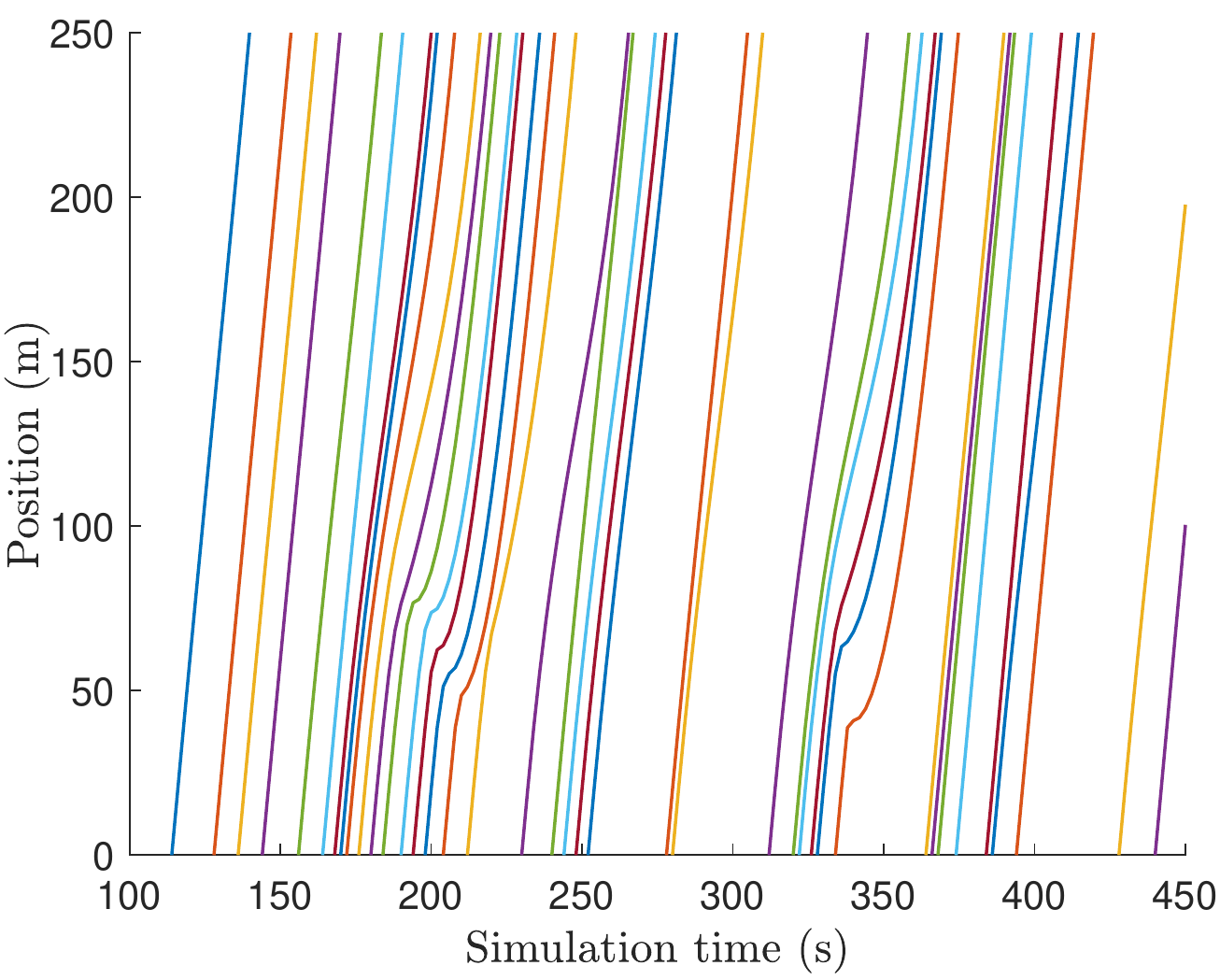}}
\subfigure[Trajectories from the DR-based strategy.]{\includegraphics[width=8cm]{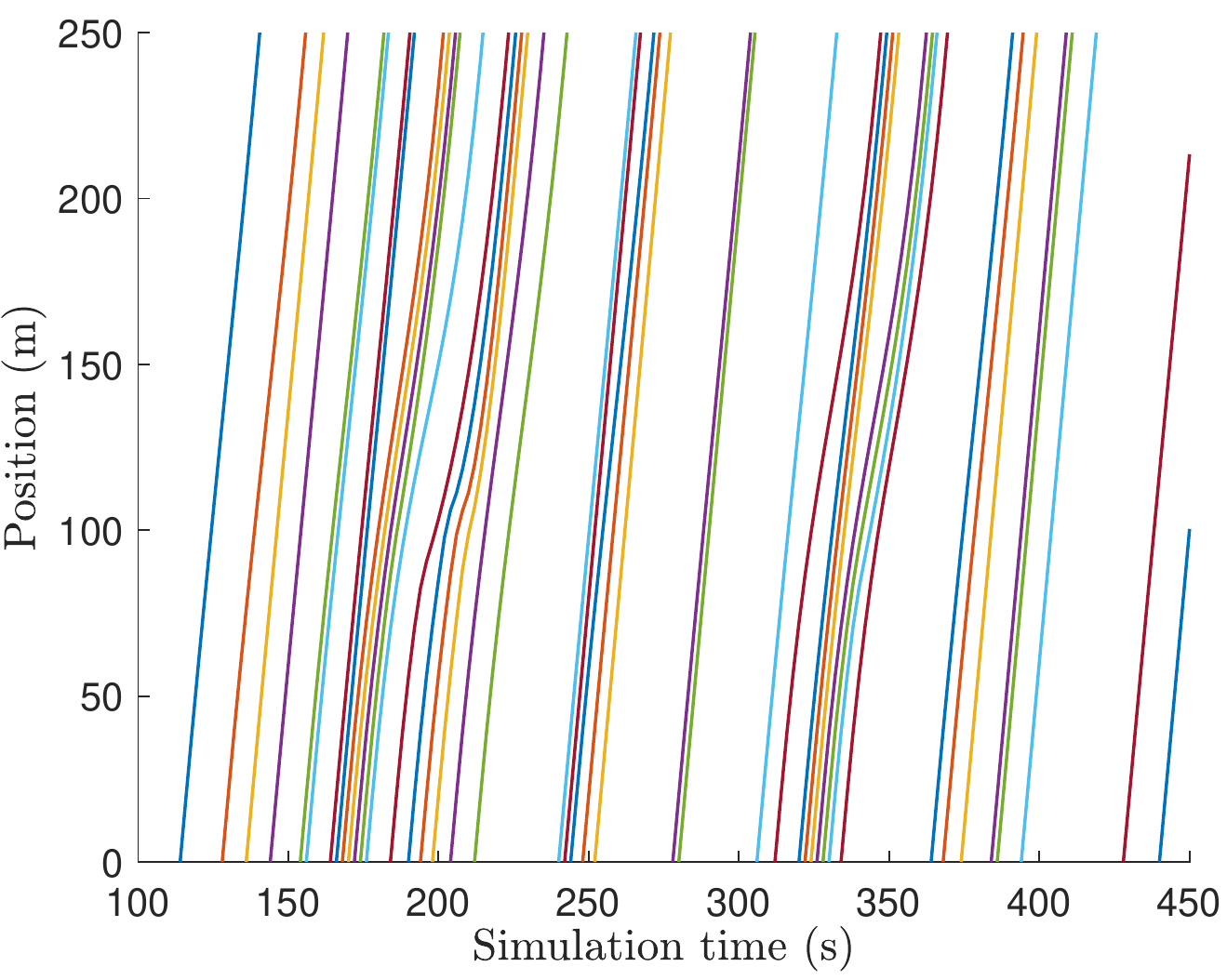}}
\caption{Partial vehicle trajectories sampled from different strategies.}%
\label{fig:stopandgo}%
\end{figure}

To validate this intuitive conclusion, Fig. \ref{fig:stopandgo} shows partial
CAV trajectories on one lane sampled from the FIFO and DR strategy,
respectively. During periods $[180,230]s$ and $[320,360]s$ we can see visible
stop-and-go activities in Fig. \ref{fig:stopandgo}(a). Due to the safety
constraints concerning the arrival time, some CAV slightly decelerates to meet
the initial constraints. However, the braking action is continuously amplified
and spreads backward. Thus, CAVs brake successively and then accelerate again,
which causes a significant added energy consumption. In contrast, the DR
strategy allows CAVs to drive more smoothly by adjusting crossing sequences.
This demonstrates how improving traffic efficiency by adjusting crossing
sequences sometimes indirectly lowers energy consumption by reducing
stop-and-go activities, especially when there is a significant gap in traffic
efficiency between the two strategies.

The performance of the modified FIFO strategy and that of the FIFO strategy
are approximately the same for this kind of intersection scenario under all
arrival rates. Although when the arrival rate is high the orders generated by
the two strategies may be a little different, the results are still similar.

\subsection{Comparison results for a geometrically asymmetrical intersection}

To explore the influence of the geometry of the intersection on the different
strategies, we set the length of one lane to be 150 $m$ while keeping the
lengths of the remaining lanes at 250 $m$. At the same time, we vary the
arrival rate from 90 veh/hour/lane to 420 veh/hour/lane to test the
performance of the four cooperative driving strategies under different traffic
demands. The results are shown in Table \ref{tab:table2}. Compared to the
results shown in Table. \ref{tab:tablel}, we can draw two conclusions.

On one hand, the most significant difference is that the modified FIFO
strategy shows a much better performance relative to the FIFO strategy.
Despite the simplicity of the idea to assign CAVs closer to the Conflict Zone
a higher priority - instead of giving CAVs that enter the Control Zone earlier
a higher priority, we obtain a significant improvement in performance when the
intersection is asymmetrical. However, the performance of the modified FIFO
strategy is still unsatisfactory when the arrival rate is high.

\begin{figure}[ptb]
\centering
\subfigure[Delay of vehicles.]{\includegraphics[width=8cm]{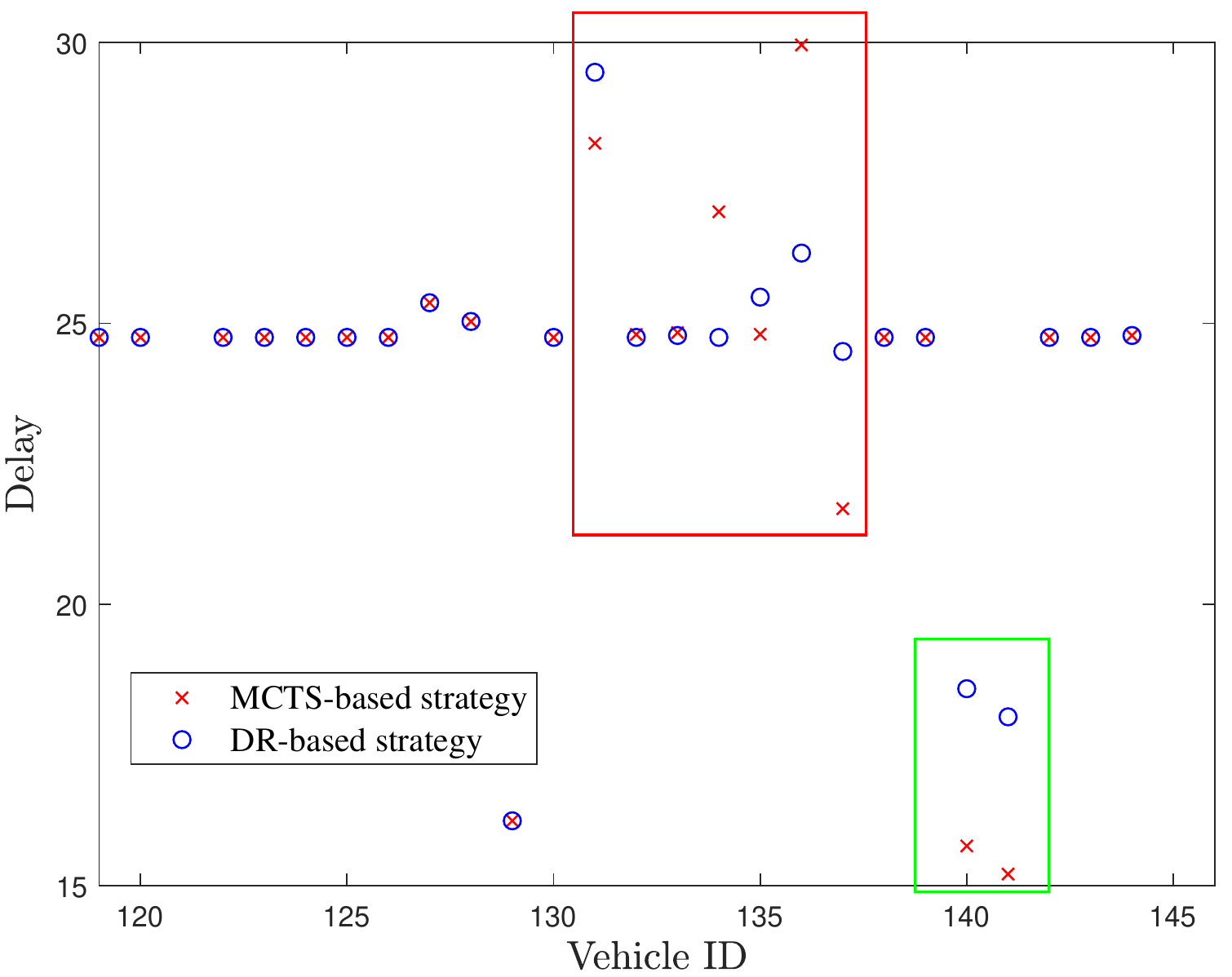}}
\subfigure[Energy consumption of vehicles]{\includegraphics[width=8cm]{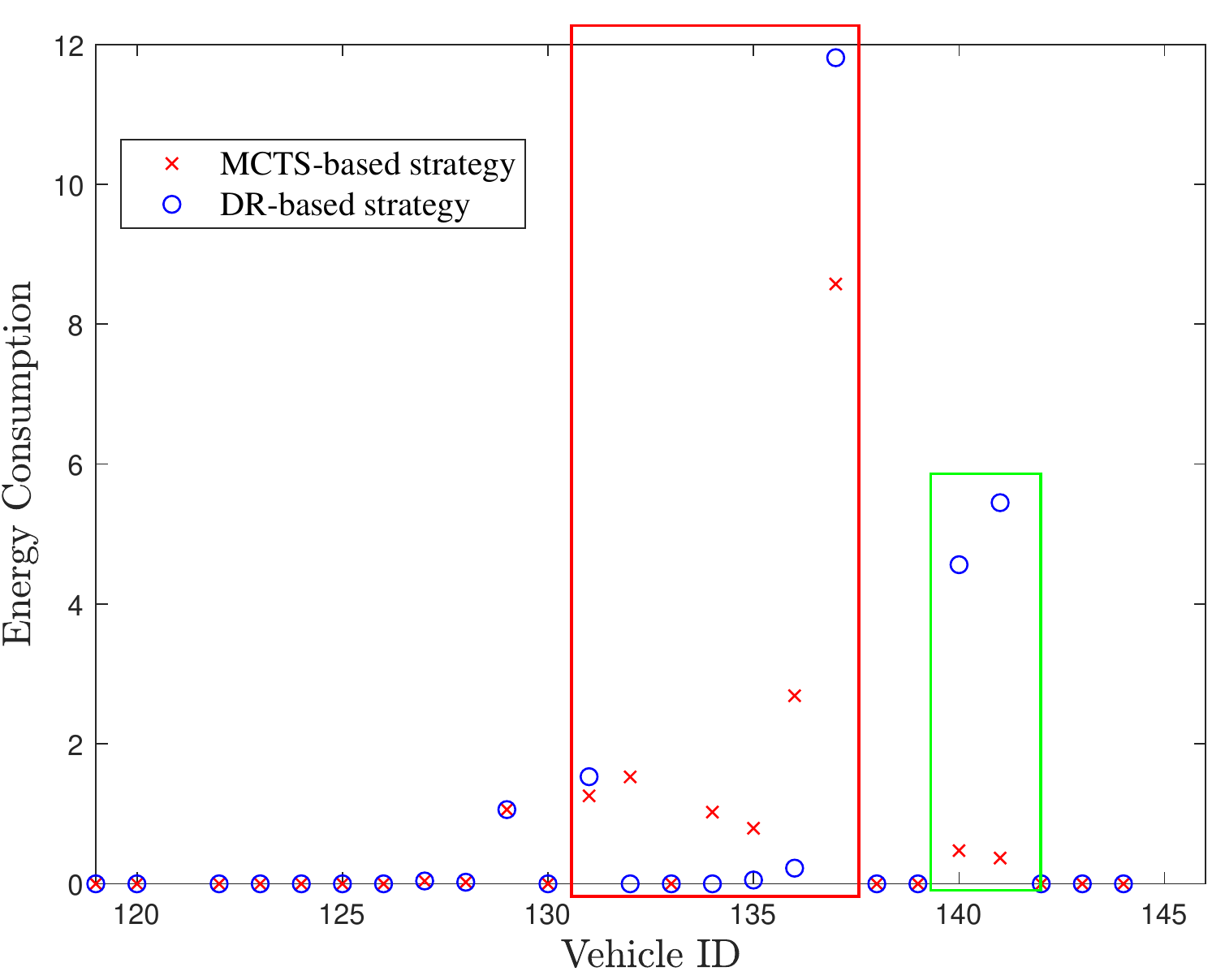}}
\caption{An example when the MCTS strategy outperforms the DR strategy in all
performance metrics.}%
\label{fig:analysis1}%
\end{figure}

On the other hand, the MCTS strategy is the best in terms of efficiency, while
the DR strategy is the best in terms of energy consumption, which is
consistent with the findings of the previous experiment. However, our results
also show that sometimes the energy consumption of the MCTS strategy is
superior to that of the DR strategy, e.g., when the arrival rate is 270
veh/h/lane. We use a snippet (the results of CAVs whose identity is from 120
to 145 as shown in Fig. \ref{fig:analysis1}) from the simulation to explain
why this happens.

\begin{table*}[ptbh]
\caption{The comparison results for the intersection with different arrival
rates}%
\label{tab:table3}
\centering
\begin{threeparttable}
\begin{tabular}{cccc}
\toprule
arrival rates (veh/h/lane) & strategies & average delay ($s$) & average energy consumption \\
\midrule
\multirow{4}[2]{*}{$\lambda_1$=180, $\lambda_2$ = 360, $\lambda_3$ = 180, $\lambda_4$ = 360} & MCTS  & 2.5275 & 0.1260 \\
& DR    & 2.5361 & 0.1107 \\
& modified FIFO & 2.9188 & 0.2076 \\
& FIFO  & 2.9188 & 0.2076 \\
\midrule
\multirow{4}[2]{*}{$\lambda_1$=225, $\lambda_2$ = 450, $\lambda_3$ = 225, $\lambda_4$ = 450} & MCTS  & 3.0165 & 0.3447 \\
& DR    & 3.0959 & 0.2503 \\
& modified FIFO & 3.7916 & 0.4511\\
& FIFO  & 3.7916 & 0.4511 \\
\midrule
\multirow{4}[2]{*}{$\lambda_1$=180, $\lambda_2$ = 270, $\lambda_3$ = 360, $\lambda_4$ = 450} & MCTS  & 2.7212 & 0.2260 \\
& DR    & 2.7793 & 0.1566 \\
& modified FIFO & 3.3132 & 0.3017 \\
& FIFO  & 3.3132 & 0.3017 \\
\midrule
\multirow{4}[2]{*}{$\lambda_1$=270, $\lambda_2$ = 360, $\lambda_3$ = 450, $\lambda_4$ = 540} & MCTS  & 3.2875 & 0.5609 \\
& DR    & 3.4612 & 0.3628 \\
& modified FIFO & 4.9855 & 0.9128 \\
& FIFO  & 4.9809 & 0.9188 \\
\bottomrule
\end{tabular}%
\end{threeparttable}
\end{table*}

As we can see from Fig. \ref{fig:analysis1}, the crossing sequences generated
by the MCTS strategy and the DR strategy for CAV 120 to CAV 130 are the same
since their performance is the same. However, they generate different crossing
sequences for CAVs 131 through 137, as shown in the red box in Fig.
\ref{fig:analysis1}(a). Since our primary goal is to decrease the average
delay, the MCTS strategy makes a large adjustment to the original crossing
sequence by forcing several CAVs ahead of CAV 137 to decelerate so as to allow
CAV 137 to pass through the Conflict Zone earlier, hence realizing a small
improvement in traffic efficiency. This improvement comes at the cost of
higher energy consumption, as can be seen in Fig. \ref{fig:analysis1}(b),
where the average energy consumption of CAVs under the MCTS strategy is higher
than that under the DR strategy. What is interesting to observe is that since
the MCTS strategy allows CAV 137 to cross first, the new coming CAVs in the
same lane (CAVs 140 and 141) have more ample road space and can access the
Conflict Zone with a higher velocity. In contrast, in the DR strategy, these
two CAVs are blocked by CAV 137, which results in both the traffic efficiency
and energy consumption of the MCTS strategy outperforming that of the DR
strategy, as shown in the green boxes in Fig. \ref{fig:analysis1}. Of course,
this is a consequence of these two CAVs randomly happening to appear.
Nonetheless, this example highlights the fact that improving traffic
efficiency sometimes indirectly decreases energy consumption.

\begin{figure}[ptb]
\centering
\subfigure[The average delay of different strategies under different arrival rates.]{\includegraphics[width=7.35cm]{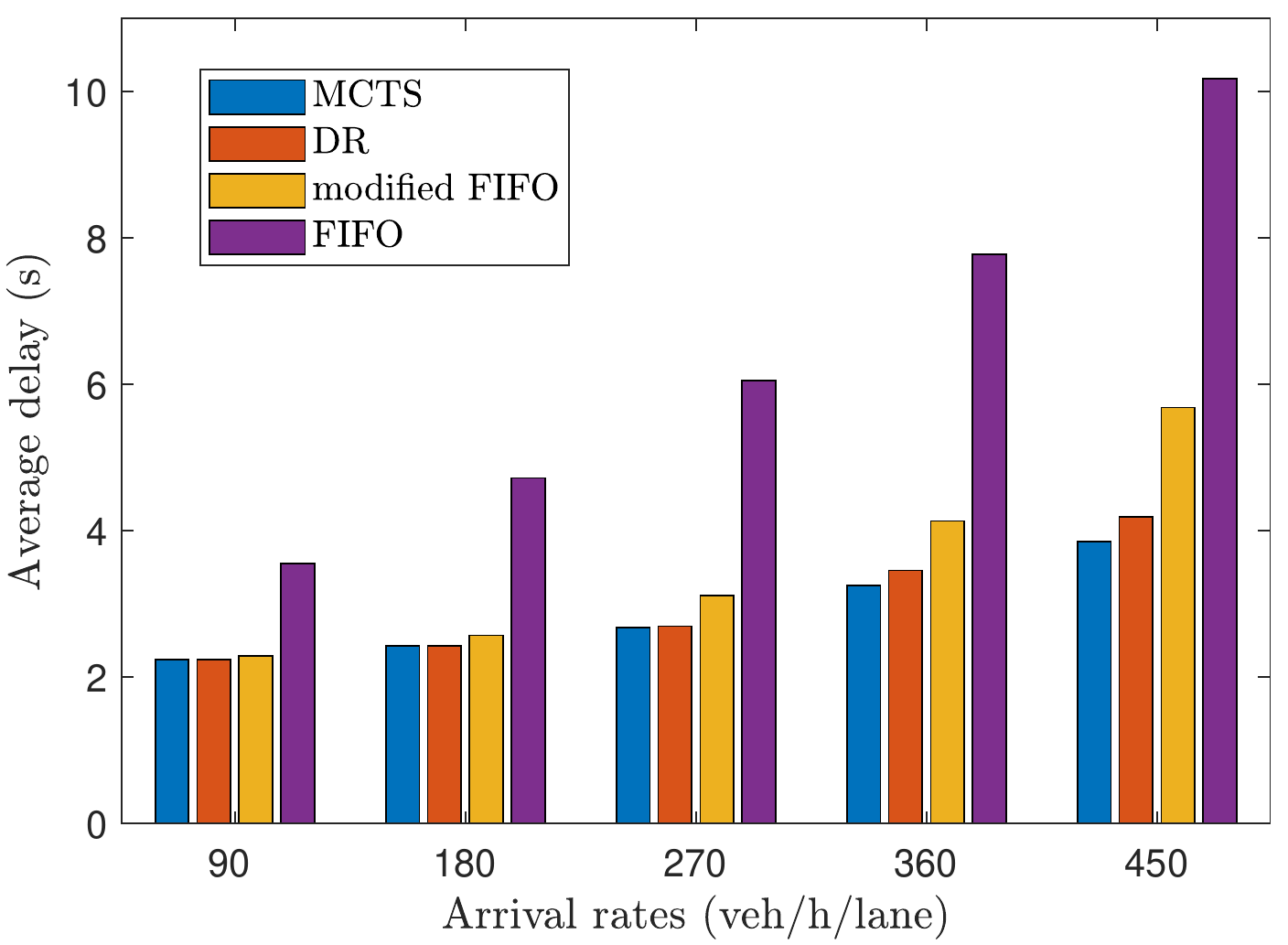}}
\subfigure[The energy consumption of different strategies under different arrival rates.]{\includegraphics[width=7.35cm]{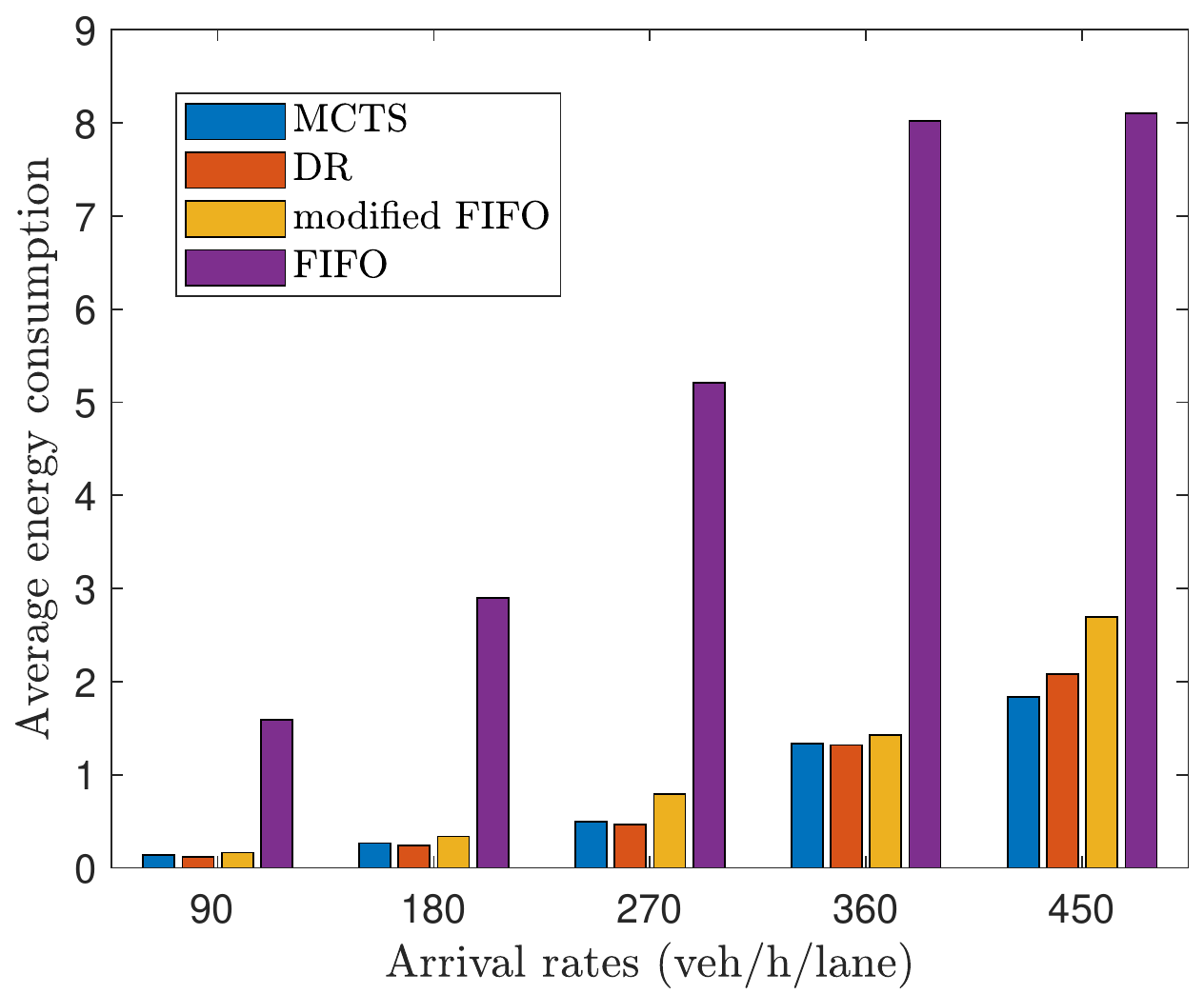}}
\caption{Comparison results of different strategies for the geometrically
asymmetrical intersection when the lanes in the EW direction are shorter than
that in the NS direction.}%
\label{fig:resulttwolanes}%
\end{figure}

We have also analyzed the situation where the lengths of lanes in the
east-to-west (EW) direction are shorter than those in the north-to-south (NS)
direction. Specifically, for the intersection shown in Fig.
\ref{fig:intersection}, the lengths of Lane 1 and Lane 3 are 150 $m$ while the
lengths of Lane 2 and Lane 4 are 250 $m$. The comparison results are shown in
Fig. \ref{fig:resulttwolanes}, and we can draw similar conclusions as in the
above situations.

\subsection{Comparison results for intersection with different arrival rates}

In this experiment, we consider a geometrically symmetrical intersection with
different arrival rates to investigate the influence of arrival rate asymmetry
on different cooperative driving strategies. We denote the arrival rates at
the entry points of Lane 1 to Lane 4 as $\lambda_{1}$, $\lambda_{2}$,
$\lambda_{3}$, and $\lambda_{4}$, respectively. The results are shown in Table
\ref{tab:table3} where we can see that they are similar to those in Table.
\ref{tab:tablel}, leading to conclusions similar to those drawn from the first
experiment. The results also suggest that the asymmetrical arrival rates do
not have a noticeable impact on the four strategies, and that the modified
FIFO strategy only outperforms the FIFO strategy in geometrically asymmetrical intersections.

\subsection{Comparison results in terms of computation time}

\begin{figure}[ptb]
\centering
\subfigure[The average computation time of different strategies in terms of the number of vehicles in the control zone.]{\includegraphics[width=7.5cm]{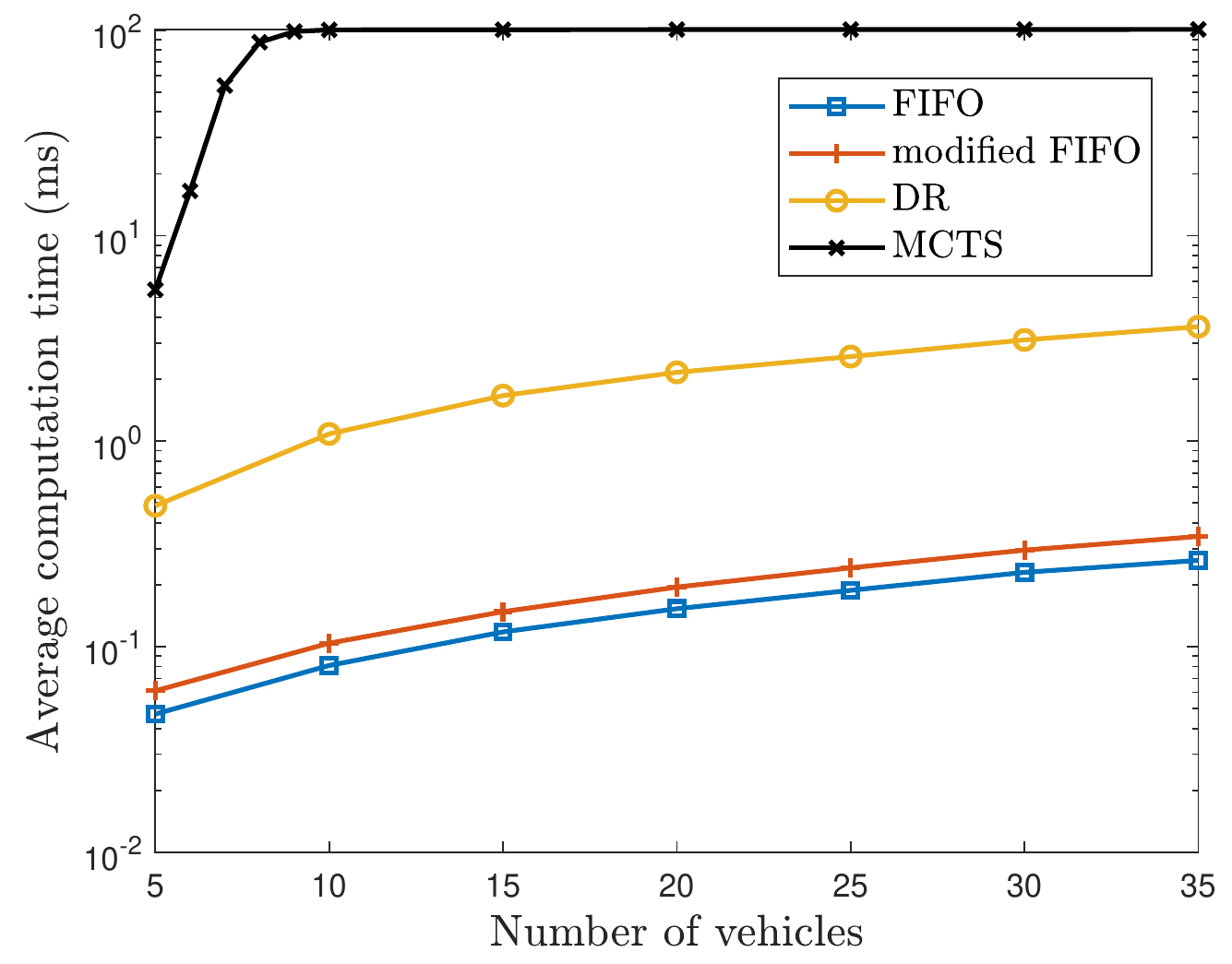}}
\subfigure[The average number of considered crossing sequences of different strategies in terms of the number of vehicles in the control zone.]{\includegraphics[width=7.5cm]{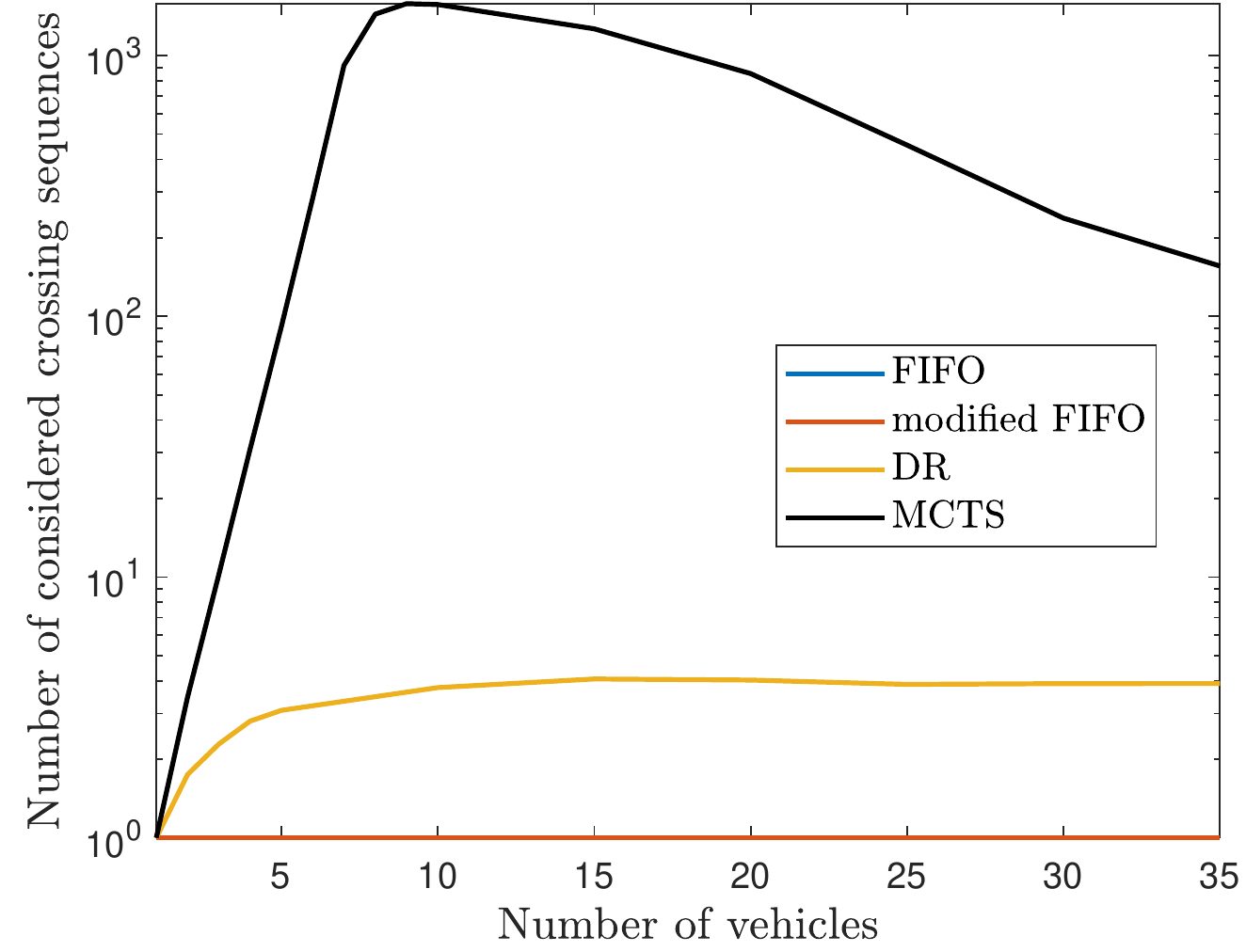}}
\caption{Comparison results of computation time and crossing sequences.}%
\label{fig:computationtime}%
\end{figure}

The computation time is vital for cooperative driving strategies to be applied
in practice. In this experiment, we study the computation time every strategy
requires to analyze the computation time and the number of crossing sequences
they have considered during that time. The results are shown in Fig.
\ref{fig:computationtime}.

As shown in Fig. \ref{fig:computationtime}(a), the computation time increases
with the number of vehicles. The FIFO strategy runs the fastest with a
computation time of less than 0.2 $ms$ even when there are as many as 35 CAVs
in the control zone. The computation time of the modified FIFO strategy is
only slightly higher. The performance of the DR strategy as a function of CAV
numbers is similar, and its computation time is smaller than 4 $ms$. In
contrast, the computation time of the MCTS strategy increases exponentially
with the number of vehicles until it reaches its assigned maximum computation
budget (which was set to 100 $ms$ in this study, which we believe to be small
enough for real-time applications). Although we can shorten or prolong the
maximum computation time, 100 $ms$ is a value that we have found to strike a
good trade-off between performance and computation.

It is of course no wonder that we can consider more crossing sequences with
more computation time. As mentioned before, the FIFO strategy and the modified
FIFO strategy only consider a single feasible crossing sequence, so their
computation time performance is similar. The number of crossing sequences that
the DR strategy considers always converges to 4 when there are enough vehicles
in the Control Zone, a fact consistent with the proof given in
\cite{zhang2018decentralized} that the expected number of crossing sequences
the DR strategy considers is equal to the number of lanes (which is 4 in our
study). For the MCTS strategy, the number of considered crossing sequences
increases exponentially with the number of vehicles at first, since the number
of feasible crossing sequences increases exponentially, and there is adequate
computation time. Then, when the number of CAVs is larger than 10, the
computation time is fixed at 100 $ms$, but the computation time for evaluating
a crossing sequence increases with the number of vehicles as the blue and red
lines show in Fig. \ref{fig:computationtime}(a). Thus, the number of
considered crossing sequences starts to decrease. However, we can still search
hundreds of feasible crossing sequences even when there is a large number of
CAVs in the Control Zone; this ensures the ability of finding a good enough
crossing sequence in practice within an acceptable computation time.

\section{The Trade-off between Fairness and Efficiency}

In this section, we explore the question regarding why resequencing still
provides benefits in heavy traffic. One might expect that in situations where
the intersection is congested, there would be little or no flexibility for
improving performance. After analyzing a large number of simulations in our
study, we believe that there are mainly two reasons for this phenomenon. The
first one is the subzone division of the Conflict Zone, and the second is that
the traffic efficiency is often improved at the cost of causing
\emph{unfairness} in extreme traffic situations.

Referring to Fig. \ref{fig:intersection}, suppose that CAV 1 in Lane 1 goes
straight, CAV 2 in Lane 2 turns left, and CAV 3 in Lane 3 goes straight, and
their distances to the conflict zone are similar. Then, it is easy to prove
that crossing sequence 132 is better than 123 since CAV 1 and CAV 3 can pass
through the intersection at the same time. Thus, even when traffic is
congested, we can still improve traffic efficiency by pairing non-conflicting
CAVs through resequencing.

When the total arrival rate at all lanes $\lambda=\sum_{i=1}^{n}\lambda_{i}$
is very close to or larger than the maximum arrival rate $\lambda_{\max}$ that
the intersection can handle (i.e., its traffic capacity), no control strategy
can alleviate congestion effectively. However, we find that strategies with
and without resequencing behave differently in this extreme situation. Suppose
that the arrival rates at all lanes are the same, i.e., $\lambda_{1}%
=\lambda_{2}=\lambda_{3}=\lambda_{4}$ and the queue lengths of all lanes are
the same, i,e., $q_{1}=q_{2}=q_{3}=q_{4}$. Then, due to the same arrival rate,
CAVs roughly arrive at all lanes evenly, which leads to CAVs in each lane
passing through the Conflict Zone in turn under the strategies without
resequencing, e.g., the FIFO strategy. However, strategies with resequencing
try to insert a new arriving CAV into a front position which provides better
performance. Sometimes, due to pairings and the randomness of the traffic
process, CAVs in one lane may leave faster than other lanes. Then, a new
arriving CAV at this lane has a much higher probability of finding a CAV to
pair with and a better position in the current crossing sequence. The
congestion in this lane may gradually dissipate while the congestion in other
lanes builds up. Thus, we conclude that strategies with resequencing tend to
block one or several lanes and allow CAVs in the remaining lanes to pair up
with non-conflicting CAVs near the Conflict Zone and pass through the
intersection quickly. Since this kind of strategy can increase the number of
vehicle pairs, it improves traffic efficiency.

\begin{figure}[th]
\centering
\includegraphics[width=8cm]{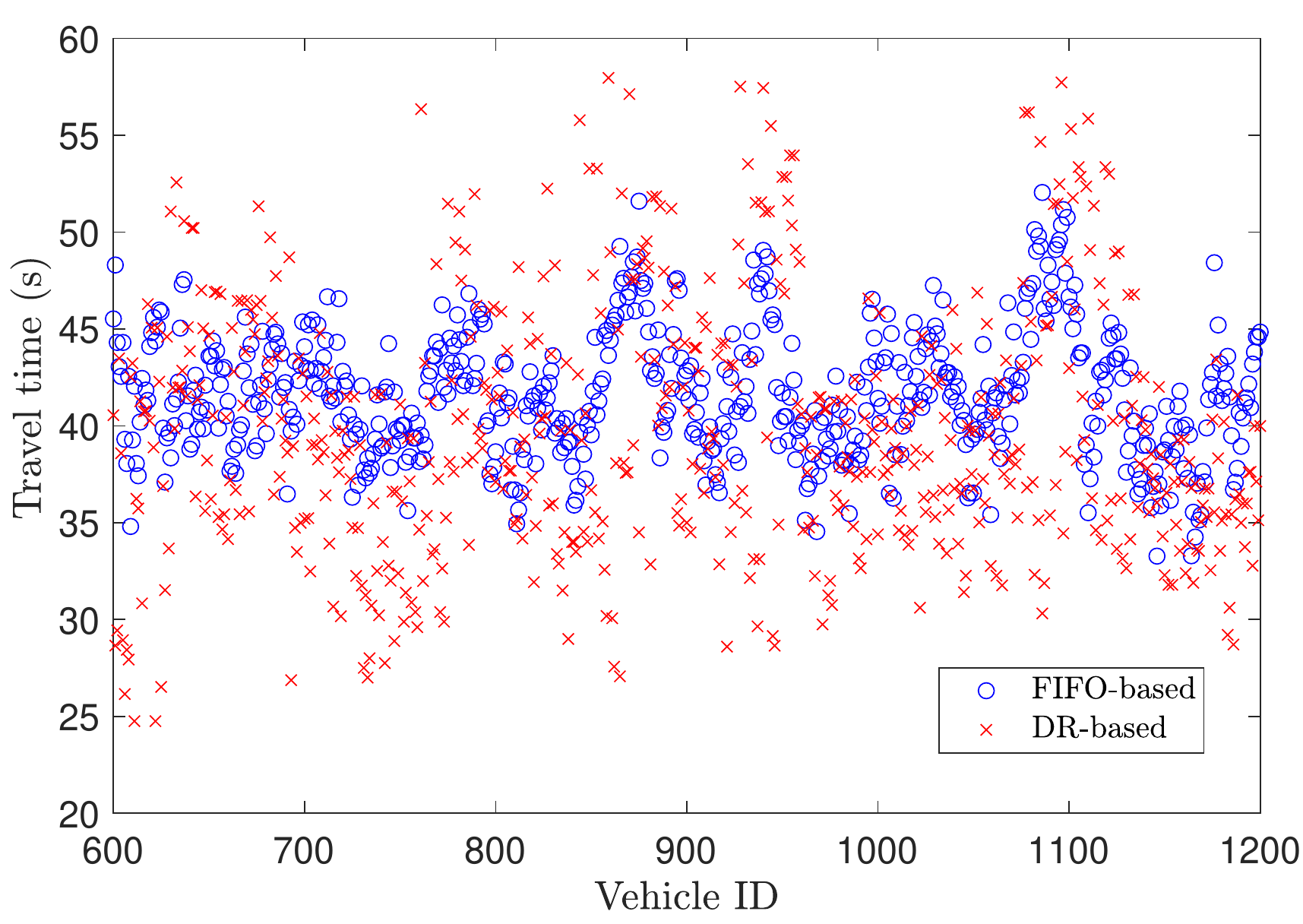}\caption{Distribution
of CAV travel times under two different cooperative strategies.}%
\label{fig:traveltimedistribution}%
\end{figure}

We use an experiment to validate this idea where we set the total arrival rate
of this intersection to a very large value ($\lambda=2$ veh/s and $\lambda
_{i}=0.5$ veh/s, $i=\{1,2,3,4\}$), much larger than the maximum arrival rate.
We also set the safety time-headway for right-turn and going straight as
$1.5s$ and for left-turn as $2.5$ s. For simplicity, we assume the safety
time-headway for all actions is $1.5$ s (the service time of the intersection
is $1.5$ s). Then, in the ideal situation, we can have two CAVs passing
through the Conflict Zone at the same time to maximize the utilization of the
road resources (that is, the intersection can serve two vehicles at one time);
clearly, the actual efficiency is lower than this. Then, in this ideal
situation, the minimum headway (service time) is $1.5/2=0.75$ s. The maximum
arrival rates should be $1/0.75=4/3$ veh/s. If the arrival rate is larger than
this value, the number of vehicles will be larger than the capacity of the
intersection leading to traffic congestion. In this case, the travel times of
CAVs under the FIFO strategy and the DR strategy are shown in Fig.
\ref{fig:traveltimedistribution}.

It is clear that the travel times of CAVs under the FIFO strategy have a much
lower variance than those under the DR strategy. In particular, the mean
travel time and standard deviation under the DR strategy are 39.06s and 6.65s,
respectively, while under the FIFO strategy they are 42.27s and 3.41s. This
example shows that traffic efficiency (lower mean travel time) may come at the
cost of unfairness and creates a natural trade-off problem. We point out,
however, that this problem typically arises at high traffic rates, since
resequencing is beneficial to \emph{all} CAVs when the arrival rate is not too
high; at high traffic rates, however, resequencing can improve the overall
traffic efficiency by sacrificing the performance of \emph{some} CAVs due to
the limited road resources to be allocated.

\begin{figure}[th]
\centering
\includegraphics[width=8cm]{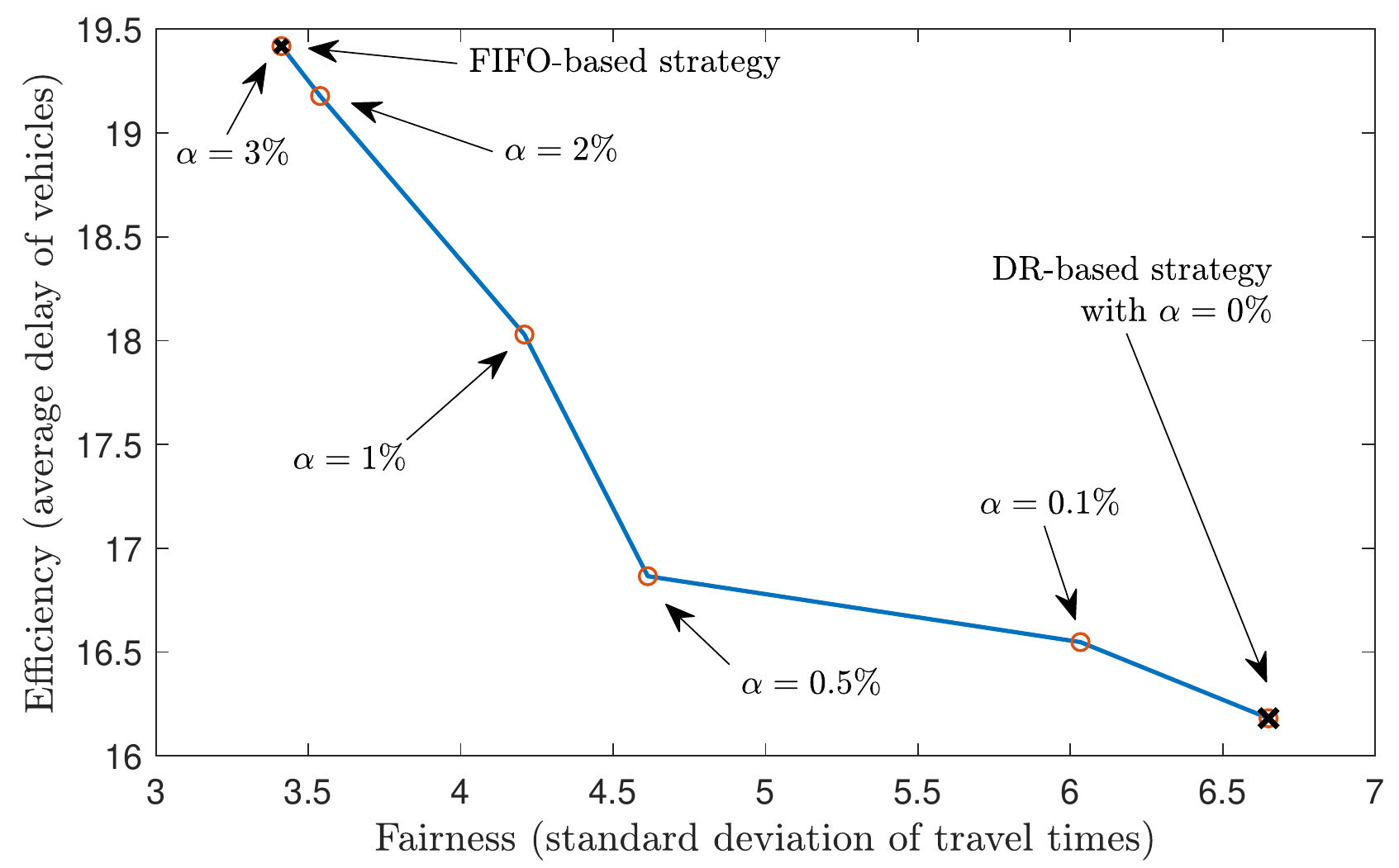}\caption{Trade-off between efficiency
and fairness under different balancing factors $\alpha$.}%
\label{fig:alpha}%
\end{figure}

A simple method of balancing performance when the DR strategy is used is based
on introducing a balancing factor $\alpha$. In particular, we only adjust the
crossing sequence when the following condition is satisfied:%

\begin{equation}
J_{new} < J_{best} - \alpha J_{new}%
\end{equation}

\noindent where $J_{new}$ is the objective value of the new crossing sequence
and $J_{best}$ is the currently optimal objective value. In other words, there
is an incentive to update the crossing sequence only when the performance of
the new crossing sequence is much better than the original one. We vary the
value of $\alpha$ from 0 to 3\% to show the trade-off between efficiency and
fairness in Fig. \ref{fig:alpha}.

Looking at Fig. \ref{fig:alpha}, it is clear that when we increase the value
of $\alpha$ we improve fairness (lower travel time standard deviation) by
decreasing efficiency, i.e., a larger $\alpha$ implies greater emphasis on
fairness. We observe that there may be a Pareto optimal point (the point with
$\alpha=0.5\%$ in Fig. \ref{fig:alpha}) that achieves a balance between the
two criteria: a perturbation to its left results in significant efficiency
relative to fairness, with the situation reversed for perturbations to its
right. This paves the way for future research in this intersting direction.

Along similar lines, for the MCTS strategy we can also make some modifications
to consider fairness in the search process. In the original MCTS strategy, we
use the following UCB1 policy to determine the most urgent expandable node:%

\begin{equation}
\mathop{\arg\max}_{i}\ Q_{i}+C\sqrt{\frac{\ln n}{n_{i}}},
\end{equation}
\noindent where $Q_{i}$ is the score of child node $i$ and $Q_{i}\in
\lbrack0,1]$. Moreover, $n$ is the number of times the current node has been
visited, $n_{i}$ is the number of times child node $i$ has been visited, and
$C$ is a weight parameter. The child node with the largest total score is
selected. The objective is to prevent significant change in order with a
resulting small benefit. Thus, we propose to add a penalty term $P_{i}$ to the
original UCB1 policy defined as%

\begin{equation}
P_{i} = \beta D_{i},
\end{equation}

\noindent where $D_{i}$ is an integer indicating how many orders are different
between the new crossing sequence and a reference crossing sequence, e.g., the
currently optimal crossing sequence or the desired crossing sequence, and
$\beta$ is a negative weight for penalizing the difference. Then, the modified
UCB1 policy is%

\begin{equation}
\mathop{\arg\max}_{i}\ Q_{i}+P_{i}+C\sqrt{\frac{\ln n}{n_{i}}}.
\end{equation}
Using this policy, the MCTS only explores significantly different crossing
sequences when it finds that such crossing sequences can bring a much improved
traffic efficiency. Note that we can also consider energy or other metrics in
the objective or modify the heuristic rules involved according to the desired
performance priorities.

\section{Conclusion and Future Research}

This paper compares the performance of some state-of-the-art cooperative
driving strategies under various influencing factors, including symmetrical
intersections, asymmetrical intersections, and asymmetrical arrival rates. Our
main conclusion is that the MCTS and DR strategies both perform well in all
scenarios and are recommended for use in practice. However, we have also
pointed out that efficiency sometimes comes at the cost of fairness to a
certain subset of CAVs. Through some modifications to these strategies, we
have shown how to control the trade-off between fairness and efficiency.

Although we have only considered an intersection with a single lane in each
direction, the conclusions of this study can be extended to other driving
scenarios, e.g., highway on-ramps and intersections with multiple lanes. There
are many problems deserve to be studied deeply, e.g., how to accelerate the
search process of the MCTS strategy further, how to choose a proper reference
crossing sequence, and so on. Due to space limitations, we will omit here and
leave it for the future.







\ifCLASSOPTIONcaptionsoff
\newpage\fi

\bibliographystyle{IEEEtran}
\bibliography{IEEEabrv,reference}

\end{document}